# scientific reports

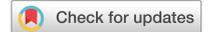

# OPEN   Application of quantum machine learning using quantum kernel algorithms on multiclass neuron M-type classification

Xavier Vasques[1,2,3✉], Hanhee Paik[4] & Laura Cif[1]

The functional characterization of different neuronal types has been a longstanding and crucial challenge. With the advent of physical quantum computers, it has become possible to apply quantum machine learning algorithms to translate theoretical research into practical solutions. Previous studies have shown the advantages of quantum algorithms on artificially generated datasets, and initial experiments with small binary classification problems have yielded comparable outcomes to classical algorithms. However, it is essential to investigate the potential quantum advantage using real-world data. To the best of our knowledge, this study is the first to propose the utilization of quantum systems to classify neuron morphologies, thereby enhancing our understanding of the performance of automatic multiclass neuron classification using quantum kernel methods. We examined the influence of feature engineering on classification accuracy and found that quantum kernel methods achieved similar performance to classical methods, with certain advantages observed in various configurations.

The field of quantum computing originated with a question posed by Feynman, namely whether it was feasible to simulate the behavior of quantum systems using a classical computer or whether a quantum computer would be required instead[1]. Since quantum algorithms were initially developed, scientists have been searching for their optimal applications and output. The first two quantum algorithms were published by Shor[2] and Grover[3], who documented that applying them to factorization and database search theoretically offers an advantage compared with classical computing. In recent years, the intersection between quantum computing and machine learning (ML) has received widespread attention and allowed the development of quantum ML algorithms[4–8]. ML and quantum computing are two technologies that may allow complex, challenging problems to be solved and progress to be accelerated in areas such as model training and pattern recognition.

Furthermore, the advances in quantum computing systems have allowed a progress in the study of quantum ML algorithms, especially with kernel methods. In classical ML, kernel methods provide a way to perform a linear classification on a highly complex data set in the framework of a support vector machine (SVM)[9–11]. Using kernel methods, quantum algorithms can be adopted through quantum kernel functions to exploit high-dimensional non-classical feature space[4,12–16] where quantum feature maps are used to encode the datapoints into inner products or amplitudes in the Hilbert space. The number of features determined the number of qubits, and a quantum circuit used to implement the feature map was of a depth that was a linear or polylogarithmic function of the dataset's size. Thus far, the studies that have been conducted to support the advantages of a quantum feature map have carefully selected synthetic datasets or applied it to small binary classification problems.

Despite the fact that research on cortical circuits has been conducted for over a century, determining how many classes of cortical neurons exist remains an ongoing and uncompleted task. Moreover, the continuous development of techniques and the availability of an increasing number of phenotype datasets have not led to the maintenance of a unique classification system that is easy to update and can consider the different defining features of neurons specific to a given type.

Neuron classification remains a topic in progress since how to designate a neuronal cell class and what the optimal features are for defining it are still questioned[17]. In this study, we demonstrate how we can apply quantum kernel methods together with classical methods for the quantitative characterization of neuronal morphologies from histological neuronal reconstructions.

[1]Laboratoire de Recherche en Neurosciences Cliniques, Montferrier-sur-Lez, France. [2]IBM Technology, Bois-Colombes, France. [3]Ecole Nationale Supérieure de Cognitique Bordeaux, Bordeaux, France. [4]IBM Quantum, IBM T J Watson Research Center, Yorktown Heights, NY 10598, USA. ✉email: xaviervasques@lrenc.org

Despite the inherent complexity and challenges that neuroscientists must deal with while addressing neuronal classification, numerous reasons exist for interest in this topic. Some brain diseases affect specific cell types. Neuron morphology studies may lead to the identification of genes to target for specific cell morphologies and the functions linked to them. A neuron undergoes different stages of development before acquiring its ultimate structure and function, which must be understood to identify new markers, marker combinations, or mediators of developmental choices. Understanding neuron morphology represents the basis of the modeling effort and the data-driven modeling approach for studying the impact of a cell's morphology on its electrical behavior and function as well as on the network dynamics that the cell belongs to.

To the best of our knowledge, the present study is the first to present a methodology to utilize quantum systems for classifying neuron morphologies with the aim of improving knowledge on the performance of the automatic multiclass classification of neurons using quantum kernel methods. We explore the feasibility of classifying neurons with quantum systems and compare classical computing, quantum simulator, and real quantum hardware. We further explore the impact of feature engineering on classification accuracy.

## Results

**Classification using feature selection.** In this study, to classify 14 classes of digitally reconstructed neuron morphologies based on the 43 morphological features described in Fig. 1, we applied several classic and quantum kernel algorithms by combining different feature rescaling and selection techniques to reduce the dataset to five features. We also considered four popular kernels to benchmark the performance of the classical SVM method (RBF, Linear, polynomial and sigmoid). Eight quantum kernel algorithms were studied. We named the first one q_kernel_zz, which applies a ZZFeatureMap and an encoding function described by Havlíček et al.,[4] who defined a feature map on n-qubits generated by the unitary. The second algorithm (q_kernel_default) applies PauliFeatureMap (paulis = ['ZI','IZ','ZZ']) with a default data mapping $\phi_S$. In addition, we used five encoding functions presented by Suzuki et al.[8] that we named q_kernel_8, q_kernel_9, q_kernel_10, q_kernel_11, q_kernel_12. The last quantum algorithm (q_kernel_training) trains a quantum kernel with quantum kernel alignment (QKA),[16] which iteratively adapts a parametrized quantum kernel to the dataset and simultaneously converges to the maximum SVM margin. In addition, selected algorithms were run by selecting 10 and 20 features. We used up to 20 qubits. The accuracy of the classification was assessed by performing cross-validation on the training dataset.

Figure 1 presents the feature's importance using XGBoost, decision tree, and random forest.

Table 1 presents the five SVMs with classical kernels that provided the best cross-validation scores when applied to sample 5. The classical algorithm that provided the best score (score = 0.91 ± 0.001) was the SVM with a radial basis function kernel and the combination of the Yeo-Johnson technique for data rescaling with the embedded decision tree classifier as a feature selection technique (selection of five features). When we used the trained model on the test dataset, the classification accuracy was 0.92. The SVM with a radial basis function used with the quantile-uniform technique for feature rescaling and an embedded decision tree classifier for feature selection also provided a cross-validation score of 0.91 ± 0.004. When we used the trained model on the test dataset, the classification accuracy was 0.91. Among all of the feature rescaling techniques, quantile transforms that map a variable's probability distribution to another probability distribution provided the optimal classification accuracies together with Yeo-Johnson. In terms of kernel functions, SVM with a radial basis function kernel provided the optimal results. The combination that provided the minimum cross-validation score was that of the quantile-uniform technique, embedded decision tree classifier, and SVM with a sigmoid kernel, which had a cross-validation score of 0.37 ± 0.01. Table 1 also provides the cross-validation scores of SVMs with quantum kernels (QSVM-Kernel classifiers) on five qubits. The algorithm that provided the best cross-validation score (0.93 ± 0.001) was q_kernel_zz with the quantile-uniform technique for data rescaling combined with the embedded decision tree classifier for feature selection (five features). This finding indicated the slightly superior precision of quantum kernel methods compared with classical ones. When we used the trained model on the test dataset, the classification accuracy was 0.93.

Furthermore, q_kernel_training associated with the quantile-Gaussian technique and embedded decision tree classifier performed better than classical top SVMs with a cross-validation score of 0.92 ± 0.001. When we used the trained model on the test dataset, the classification accuracy was 0.92. In terms of quantum kernel methods, q_kernel_training and q_kernel_zz provided the most accurate results among the quantum algorithms. The combination that provided the minimum cross-validation score was that of MaxAbsScaler, XGBoost, and q_kernel_10, which had a cross-validation score of 0.55 ± 0.005.

Figure 2 presents the cross-validation scores of q_kernel_training run on five qubits with a combination of the quantile-Gaussian technique for data rescaling and a decision tree for feature selection; classical SVM with RBF kernel using the combination of Yeo-Johnson for data rescaling and a decision tree for feature selection; and q_kernel_zz run on five qubits using the combination of the quantile-uniform technique and a decision tree, which were applied to the samples described in Fig. 1. We observed that the three algorithms behaved the same as classical SVM, performing better on samples 1 to 4. Both q_kernel_zz and q_kernel_training performed slightly better than classical SVM on sample 5, indicating an accuracy improvement over the amount of data, such as in classical SVM.

To compare quantum and classical algorithms with the same feature rescaling and feature selection method, we combined the quantile-uniform technique for feature rescaling and decision tree for feature selection with q_kernel_zz on five qubits as well as classical SVMs with RBF, linear, polynomial, and sigmoid kernels to classify neuron morphologies using five selected features on samples 1–5 (Fig. 2). The q_kernel_zz performed better than the SVM with sigmoid on all samples as well as the SVM with linear kernels on samples 1, 4, and 5 (the score was the same on sample 3). The SVM algorithms with a polynomial and RBF kernels performed better on all samples except for sample 5. In addition, we combined the quantile-Gaussian technique for feature rescaling and a decision tree for feature selection to compare the performance of q_kernel_training (five qubits) and classical SVMs with RBF, linear, polynomial, and sigmoid kernels (Fig. 2). The q_kernel_training performed better than the SVM with sigmoid and polynomial kernels on all samples. Furthermore, the SVM algorithms with RBF and linear kernels performed better than q_kernel_training on all samples except for sample 5.

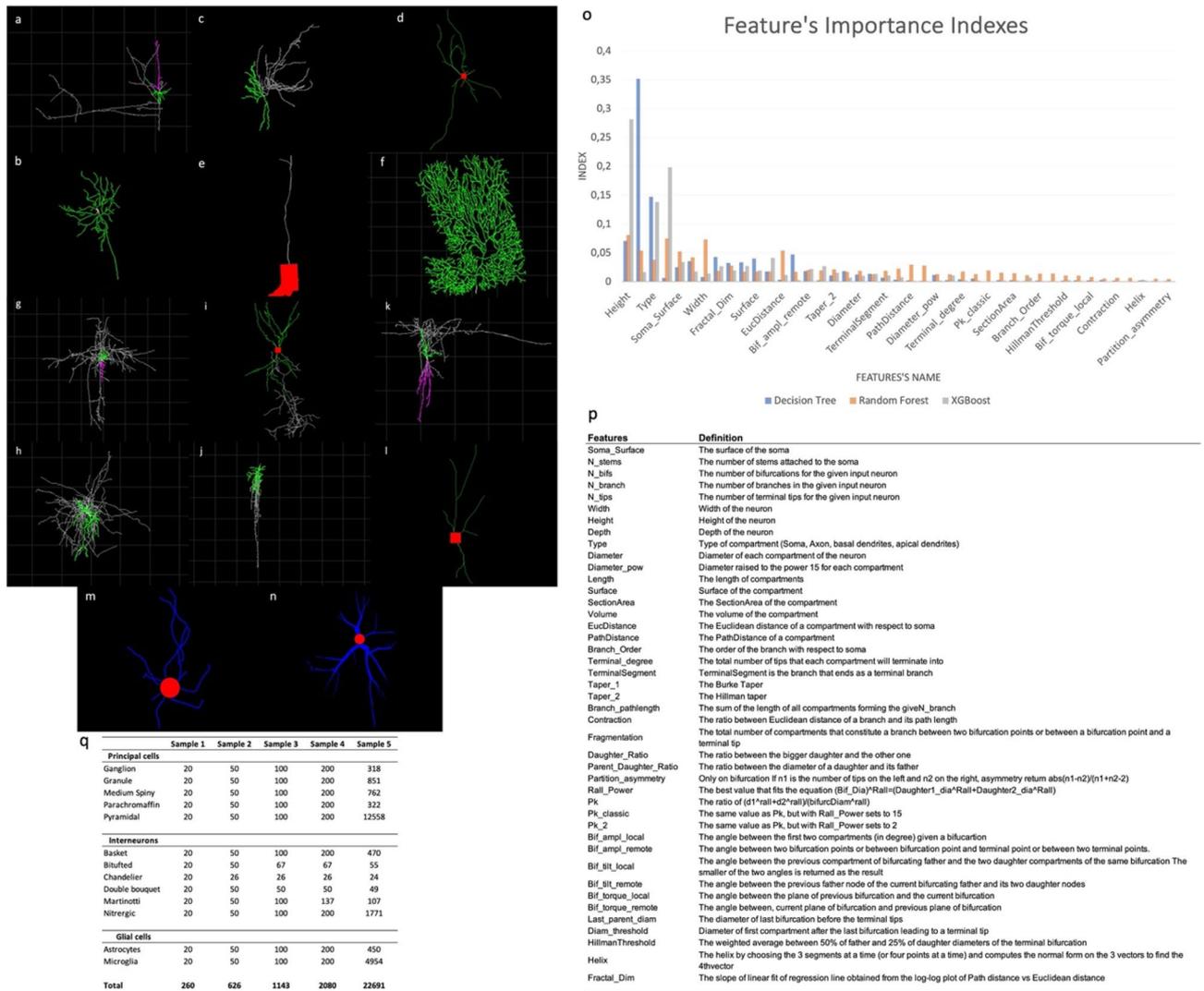

**Figure 1.** Neuronal morphologies[44–47] from the NeuroMorpho-rat dataset. Principal neurons are presented, such as (**a**) a pyramidal cell (layer 4, C010398B-P2) from the rat somatosensory neocortex; (**b**) a ganglion cell (LY8-RGC1) from the retina; (**c**) a granule cell (03D23APV-1) from the hippocampus; (**d**) a medium spiny cell (1-1-DE) from the nucleus accumbens; (**e**) a parachromatin cell (D20c) from the adrenal medulla; and (**f**) a Purkinje cell (alxP) from the cerebellum. Interneurons from the rat somatosensory neocortex are presented, such as (**g**) a basket cell (layer 2–3, C010398B-I4); (**h**) a bitufted cell (layer 4, C020600C1); (**i**) a chandelier cell (layer 2–3, C231001B2); (**j**) a double bouquet (layer 2–3, C060400B2); (**k**) a Martinotti cell (layer 2–3, C050398B-IA); and (**l**) a nitrergic cell (layer 5–6, RatS1-1-1). In addition, we demonstrated two types of glial cells: (**m**) a microglia cell (farsight624) from the frontal neocortex and (**n**) astrocyte cells (A1-CA1-L-C63x1zACR1) from the hippocampus. (**o**) Feature's importance using XGBoost, decision tree, and random forest applied to the entire dataset (sample 5). (**p**) Definition of the 43 morphological features extracted for each neuron. They were extracted using the L-Measure tool,[40] providing quantitative morphological measurement from neuronal reconstruction (http://cng.gmu.edu:8080/Lm/help/index.htm). (**q**) Dataset with the number of neuron morphologies for multiclass classification. From the 27,881 extracted neurons, 22,691 neurons (sample 5) remained after the application of Mahalanobis distance transformation and the suppression of all neurons with a soma surface equal to 0.

Figure 2 presents the cross-validation scores of q_kernel_zz (with the quantile-uniform technique for data rescaling and decision tree for feature selection), SVM RBF (with the quantile-uniform technique for data rescaling and decision tree for feature selection), and SVM RBF (with Yeo-Johnson for data rescaling and decision tree for feature selection) using 20 as well as 10 selected features, which were applied to all of the samples described in Fig. 1. For q_kernel_zz, we used 20 and then 10 qubits. For 20 qubits, we observed the progression of q_kernel_zz's performance across samples to be much more pronounced compared with the classical SVMs. The cross-validation score of q_kernel_zz progressed to 0.27 between samples 1 and 2, 0.2 between samples 2 and 3, 0.09 between samples 3 and 4, and 0.15 between samples 4 and 5, reaching an equivalent cross-validation score to those of classical algorithms. For 10 qubits, as we already observed for five selected features, q_kernel_zz's performance started behind that of the classical SVM RBF, improving across samples until it reached a similar score on sample 5. Both quantum and classical algorithms slightly improved the classification score with the increase in the number of selected features.

| Rescaling | Feature extraction | Algorithms | Cross-validation mean | Cross-validation standard deviation |
|---|---|---|---|---|
| Support vector machine with classical kernel functions | | | | |
| *Feature extraction techniques applied to reduce sample 5 dataset to 2 features* | | | | |
| Yeo-Johnson | LDA | SVM_rbf | 0.89 | 0.002 |
| Yeo-Johnson | LDA | SVM_linear | 0.89 | 0.003 |
| Yeo-Johnson | LDA | SVM_poly | 0.88 | 0.002 |
| Quantile-Gaussian | LDA | SVM_rbf | 0.87 | 0.001 |
| Quantile-Uniform | LDA | SVM_linear | 0.86 | 0.001 |
| Support vector machine with quantum kernel functions | | | | |
| *Feature extraction techniques applied to reduce sample 5 dataset to 2 features running on 2 qubits* | | | | |
| MaxAbsScaler | LDA | q_kernel_zz | 0.75 | 0.005 |
| RobustScaler | LDA | q_kernel_zz | 0.75 | 0.005 |
| MinMaxScaler | PCA | q_kernel_zz | 0.73 | 0.005 |
| MaxAbsScaler | PCA | q_kernel_zz | 0.73 | 0.005 |
| MaxAbsScaler | TruncatedSVD | q_kernel_zz | 0.71 | 0.006 |
| Support vector machine with classical kernel functions | | | | |
| *Feature selection techniques applied to reduce sample 5 dataset to five features* | | | | |
| Yeo-Johnson | embedded_decision_tree_classifier | SVM_rbf | 0.91 | 0.001 |
| Quantile-Uniform | embedded_decision_tree_classifier | SVM_rbf | 0.91 | 0.004 |
| Quantile-Gaussian | embedded_xgboost_classification | SVM_rbf | 0.90 | 0.002 |
| Yeo-Johnson | embedded_decision_tree_classifier | SVM_poly | 0.90 | 0.003 |
| Quantile-Gaussian | embedded_decision_tree_classifier | SVM_rbf | 0.90 | 0.003 |
| Support vector machine with quantum kernel functions | | | | |
| *Feature selection techniques applied to reduce sample 5 dataset to five features running on 5 qubits* | | | | |
| Quantile-Uniform | embedded_decision_tree_classifier | q_kernel_zz | 0.93 | 0.001 |
| Quantile-Gaussian | embedded_decision_tree_classifier | q_kernel_training | 0.92 | 0.002 |
| Quantile-Uniform | embedded_xgboost_classification | q_kernel_zz | 0.9 | 0.003 |
| StandardScaler | embedded_xgboost_classification | q_kernel_zz | 0.89 | 0.003 |
| StandardScaler | embedded_decision_tree_classifier | q_kernel_zz | 0.88 | 0.005 |

**Table 1.** Five-fold cross-validation scores from using SVM with classical and quantum kernels. A combination of feature rescaling and feature selection techniques were applied to reduce the number of features to five running on 5 qubits. A combination of the feature rescaling and feature extraction techniques (number of components = 2 running on 2 qbits) was applied to reduce the number of dimensions. The different algorithms were applied to the full dataset (sample 5). We present the top five algorithms that provided the best accuracy.

### Neuron morphology classification using feature extraction.

After rescaling data with each of the techniques described in the method section, we combined the outputs with feature extraction techniques to dimensionally reduce the number of features of the dataset to two principal components.

Table 1 presents the SVMs' cross-validation scores with classical and quantum (two-qubit) kernels. The results from applying the algorithms on sample 5 are displayed. We present the five algorithms that provided the best cross-validation accuracy scores. Here, the five algorithms that provided the optimal accuracy were SVMs with classical feature maps, which indicates that feature extraction techniques may not be the most suitable techniques for the quantum kernel methods used in this study. SVM RBF with Yeo-Johnson for data rescaling and LDA for feature extraction had a cross-validation score of $0.89 \pm 0.002$. By contrast, the best quantum algorithms (q_kernel_zz), which used MaxAbsScaler for data rescaling and LDA for feature extraction, had a cross-validation score of $0.75 \pm 0.005$. When we examined the application of SVM RBF with Yeo-Johnson and LDA to the different samples, we observed a cross-validation score of 0.73 for sample 4, 0.68 for sample 3, 0.59 for sample 2, and 0.48 for sample 1. By contrast, the best quantum algorithm applied to sample 4 was q_kernel_zz with MinMaxScaler and PCA with a score of 0.56. The best cross-validation scores were provided by q_kernel_zz with MaxAbsScaler and PCA for sample 3 with a score of 0.51, by q_kernel_zz with MaxAbsScaler and PCA for sample 2 with a score of 0.56, and by q_kernel_zz with Yeo-Johnson and ICA without PCA for sample 1 with a score of 0.54.

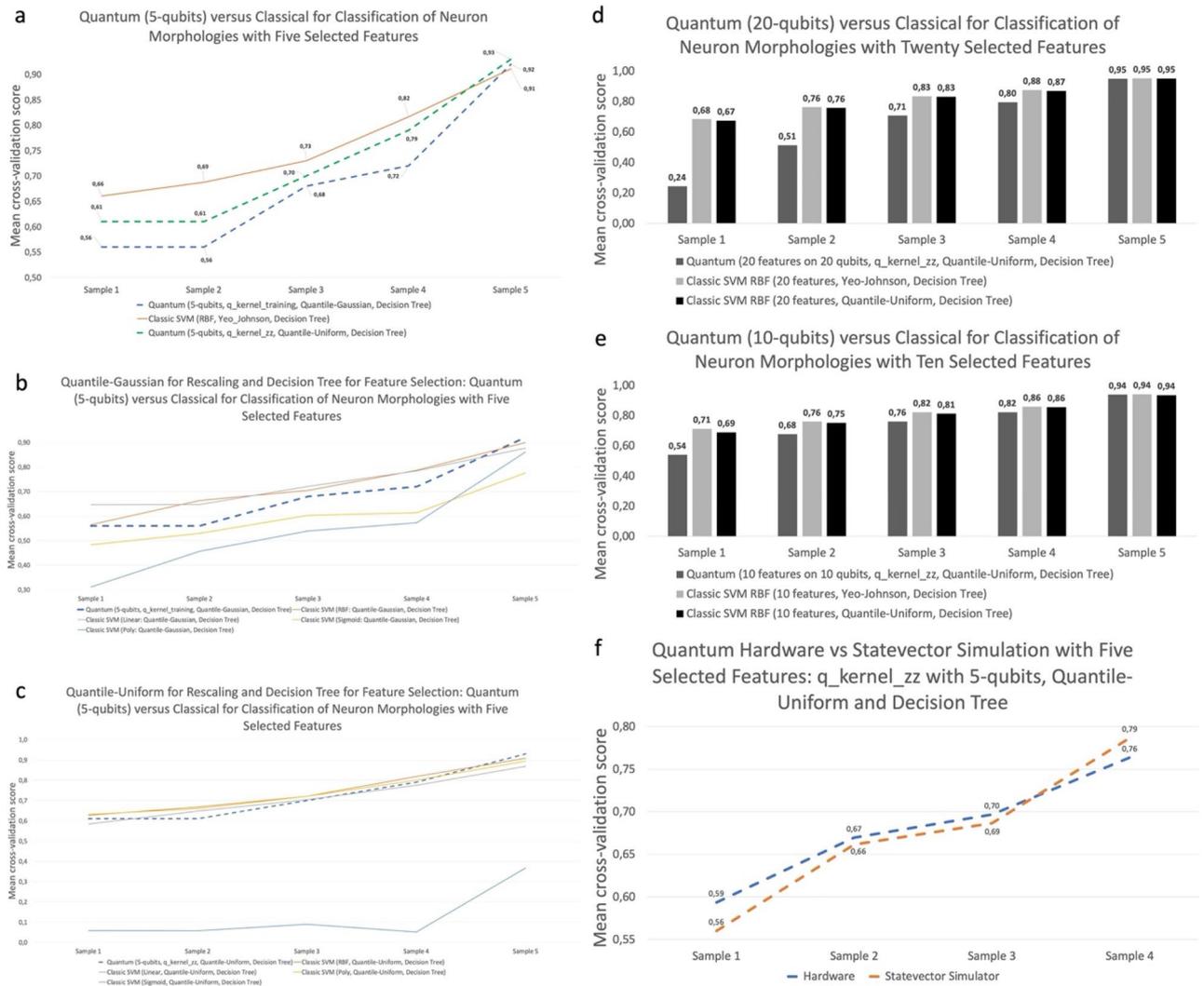

**Figure 2.** Cross-validation scores of the different algotihms (**a**) Cross-validation scores of q_kernel_training run on five qubits with the combination of the quantile-Gaussian technique for data rescaling and a decision tree for feature selection; classical SVM with RBF kernel using the combination of Yeo–Johnson for data rescaling and a random forest for feature selection; and q_kernel_zz run on five qubits using the combination of the quantile-uniform technique and a decision tree, which were applied to the samples described in Fig. 1. The results were obtained by running the algorithms with five selected features. (**b**) Five-fold cross-validation scores obtained with the combination of the quantile-uniform technique for feature rescaling and a decision tree for feature selection to compare the performance of q_kernel_zz (five qubits) and classical SVMs with RBF, linear, polynomial, and sigmoid using five selected features on all of the samples described in Fig. 1 (sample 1 = 260 neurons, sample 2 = 626, sample 3 = 1143, sample 4 = 2080, sample 5 = 22,691). (**c**) Five-fold cross-validation scores obtained with a combination of the quantile-Gaussian technique for feature rescaling and a decision tree for feature selection to compare the performance of q_kernel_training (five qubits) and classical SVMs with RBF, linear, polynomial, and sigmoid kernels using five selected features on all samples. (**d**) Five-fold cross-validation scores obtained from the combination of the quantile-uniform and Yeo–Johnson techniques for feature rescaling and a decision tree for feature selection, comparing the performance of q_kernel_zz (20 qubits) and classical SVMs with RBF to classify neuron morphologies using 20 selected features on all of the samples described in Fig. 1 (sample 1 = 260 neurons, sample 2 = 626, sample 3 = 1143, sample 4 = 2080, sample 5 = 22,691). (**e**) Five-fold cross-validation scores obtained from a combination of the quantile-uniform and Yeo–Johnson techniques for feature rescaling and a decision tree for feature selection to compare the performance of q_kernel_zz (10 qubits) and classical SVMs with RBF to classify neuron morphologies using 10 selected features on all samples. (**f**) Cross-validation scores of q_kernel_zz run with five qubits with a combination of the quantile-uniform technique for data rescaling and a decision tree for feature selection on both quantum hardware and statevector simulation applied to the different samples described in Fig. 1.

**Results from quantum computer hardware.** Addressing quantum ML performance on today's noisy quantum computer hardware is of major interest. For neuron morphology classification, we used the quantum algorithm that provided the best cross-validation score on the StatevectorSimulator, namely the q_kernel_zz (five qubits for five selected features) with a combination of the quantile-uniform technique for data rescaling and the embedded decision tree classifier for feature selection. Four 27-qubit superconducting quantum computers available on the IBM Quantum Services were used to run the quantum algorithms (Fig. 3). Due to the limited access time and the running time, we performed four runs by applying the algorithms to samples 1–4 with 5 features selected (five qubits). In the future, the running time is expected to significantly reduce with hardware improvements. For every kernel entry, we used 1024 shots to reduce the statistical uncertainties in evaluating kernel entries on quantum hardware. As indicated by our results in Fig. 2, the difference in the performance achieved by the quantum computer hardware and a quantum computer simulator can be explained by the effect of quantum hardware noise and fluctuated among the hardware runs[18].

## Discussion

To understand how the brain works, the functional identification of distinct types of neurons has been and remains a crucial challenge. For over half a century, neuronal classification based on morphology alone remained a dominant theme in neurobiology; however, it fell out of fashion as physiological and molecular methods matured, and mechanistic, 'hypothesis-driven' research came to be valued over projects with 'merely descriptive' aims[19]. Nevertheless, the first step in neuron taxonomy remains identifying the morphology of neurons, since they are a direct reflection of their synaptic connections, followed by understanding what they do. Identifying the different types of neurons is not only critical for understanding the microscopic anatomy of the nervous system. The different shapes and sizes of the various neuronal types are the expression of a fundamental—probably developmental—relationship among the neurons of a specific type[20]. Furthermore, many disease processes that affect the nervous system exhibit selectivity and may only involve certain neuronal types while sparing others. To a larger extent, the shape and extent of the dendritic tree define the neuron's role in the organization of the nervous system as a whole, since the neuron determines the field from which it receives its input. Dendrites form relatively late in the development of neurons, and always preceded by the outgrowth of axons and axon connection. The high-throughput generation of data is expected to enable the learning of a systematic taxonomy from data by considering molecular, morphological, and electrophysiological features[21]; however, this huge amount of data requires appropriate storage and processing. In a transition in neuroscience over the last decade, data scientists in computational modeling and database experts have gained a major place in the field by extracting knowledge and finding regularity within the data. A central role is given to a proper digital infrastructure and the technical requirements necessary for storing, representing, and analyzing the expected deluge of data necessary for cracking the neural code[22]. Digital datasets in nearly all subfields of neuroscience require the right computers to be deeply integrated in daily practice. Moreover, a detailed characterization of cellular anatomy is essential for elucidating neuronal computation. Digital reconstructions increase the reliability of anatomical quantifications and enable biologically realistic computational simulations for investigating the neuronal structure–activity relationship[23].

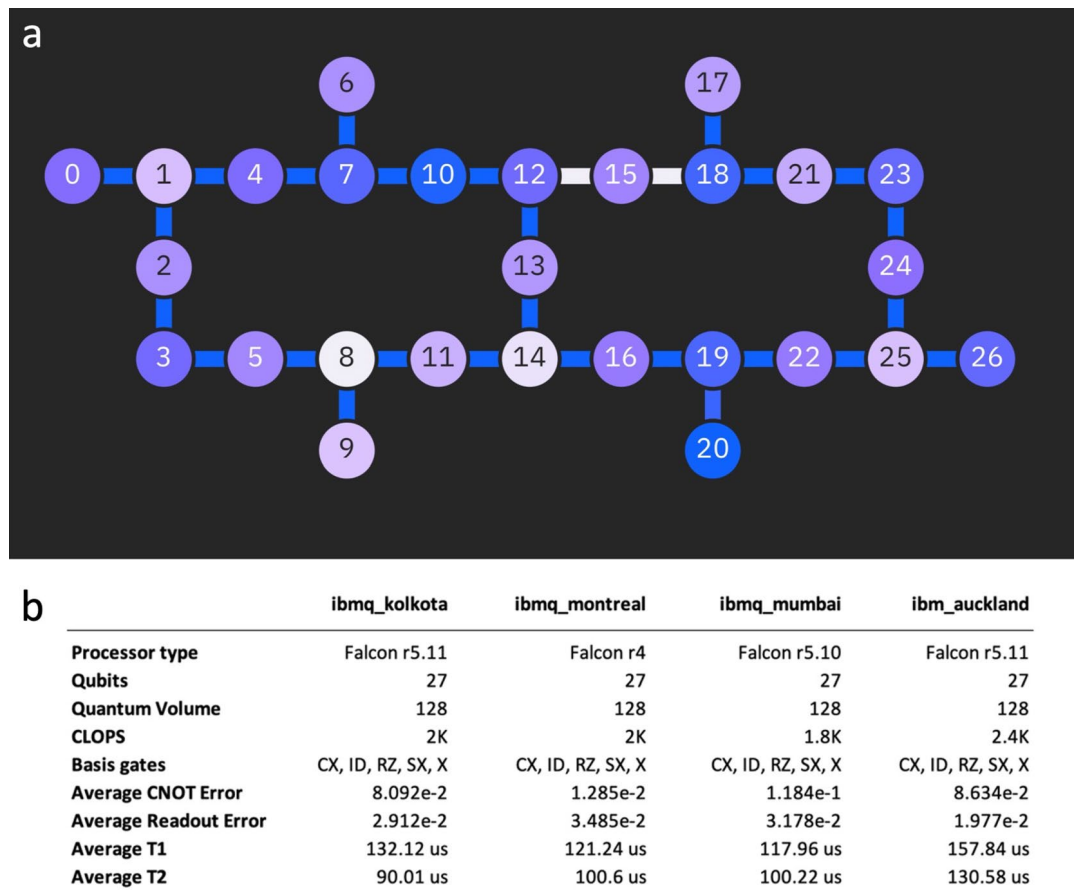

**Figure 3.** Quantum Systems (**a**) Qubit connectivity of the four 27-qubit superconducting quantum computers. Lighter colors mean a higher T2 time for qubits and lower fidelity for coupling. (**b**) Characteristics of the four quantum systems.

Many improvements have been achieved in neural classification based on morphology, molecules, electrophysiology, transcriptomes, genomes, and biophysics, as well as for better understanding biological structures and functions both systematically and reproducibly[24]. Connectivity-based neuronal classification strategies can be integrated with other cell type–specific information, such as molecular identities (mRNA expression, epigenomics, or genetic labeling), electrophysiological properties, and functional specificities. Single-cell transcriptomic signatures can be associated with specific differential electrophysiological and axon projection properties[25,26]. Morphological diversity also relates to the intrinsic functional differences between neuron classes[27]. Neuronal morphology affects network connectivity, plasticity, and information processing. Each neuron must have proper synaptic partners to function effectively and accurately[28] and also for establishing the proper circuitry. Distinct neuronal cell types acquire and maintain their identity by expressing different genes. Categorizing cortical neurons into types and then studying the roles of the different types in the function of the circuit represent an essential step toward understanding how different cortical circuits produce distinct computations[19]. The probing of single-cell mRNA opens up new approaches for understanding brain circuitry, plasticity, and pathology, thus refining the concept of the neuronal cell type[29]. The increasing number of neuronal features progressively complete cell types. Computational approaches are necessary for quantifying the intricate relationship between neuronal morphology (structure) and physiology (activity)[28,30].

More precise knowledge of the identity of the types of neurons that constitute the neural circuits found in different species may serve and benefit the field of comparative connectomics. When achieved, neuron classification will enable genetic access to specific neurons so that they can be marked, manipulated, and studied in diseases that involve specific cell types. A unifying definition of neuronal type should involve multiple criteria (i.e., physiological, morphological, and molecular properties). This implies that they co-vary,[19] rather than being a single 'essential' feature,[31] and also that the criteria for defining types should be quantitative.

Moreover, studying and clarifying neuronal morphology will allow progress in the knowledge of both molecular and physiological properties. Recordings from and morphological reconstructions of thousands of neurons from cortical brain slices have been used to classify them into hundreds of morpho-electrical types[32]. Here, we questioned whether quantum computers may provide the same output as classic computers for the morphological classification of neurons. The most mature, scalable, and useful technique for the molecular profiling of cell type diversity is single-cell RNA sequencing (scRNA-seq). This method is easily amenable to automation and, if applied at a sufficiently large scale, could drive a first "complete" cell-type classification—a potential further application of our approach using quantum computers. By combining classical kernel methods and quantum models, quantum kernel methods such as QSVM can shape new approaches in ML[4,12–15] by using the quantum state space as the feature space. From these algorithms, we expect to obtain an acceleration and increased accuracy on complex data, such as multiple-criteria neuron type classification. However, to fully exploit the computational advantage from quantum kernels, it is necessary to understand the underlying data structure to identify an optimal feature map[4,15,18].

Liu and colleagues[13] highlighted that quantum machine learning algorithms are capable of providing a quantum computing speed-up without assuming quantum access to data. They succeeded in attaining a substantial quantum advantage by utilizing a quantum version of a conventional support vector machine, which estimates a kernel function aided by a fault-tolerant quantum computer. The process of mapping data samples to a quantum feature space, where kernel entries can be estimated as the transitaion amplitude of a quantum circuit, bears particular relevance to our research. Similar to Liu et al., we utilize quantum kernel estimation algorithms in our methodology. However, beyond the potential speed-up, these algorithms provide the opportunity of encoding high-dimensional data that may confer a classification advantage. Given the complexity and high dimensionality of neuronal data, our methodology could outperform classical methods in more accurately classifying neuron types. We anticipate these algorithms to expedite processing and enhance accuracy on complex data, such as multi-criteria neuron type classification (morphology, electrophysiology, molecules, transcriptome, genome, biophysics), facilitating more systematic and reproducible understanding of biological structures and functions. In our study, we demonstrated the application of quantum kernel methods in classifying neuronal morphologies using quantum algorithms. We trained both classical and quantum kernel methods with varying kernels and empirically explored different combinations of feature rescaling, feature extraction, feature selection, algorithms, and counts of features and qubits to assess potential advantages of quantum kernels on our datasets. In the present study, we have successfully demonstrated a methodology how we can apply the quantum kernel methods for classifying neuron morphologies on quantum computers. The work of Glick et al.[16] introduces a unique class of quantum kernels constructed for data with group structures. These quantum kernels are estimated via unitary circuits and potentially offer a quantum advantage, given certain computational hardness assumptions. Their research marks a substantial advancement in the identification of quantum kernels that provide an advantage in the classification of real-world data. Considering the potential interpretation of neuron morphology data as having a mathematical group structure (with different M-types of neurons being regarded as distinct groups), the covariant quantum kernel approach proposed by Glick et al. could prove applicable and advantageous within our framework. The incorporation of these methods and their application to broader datasets could enhance the efficacy of our approach in extracting essential features from the data, resulting in more accurate and efficient neuron type classification.

In this demonstration, we trained classical and quantum kernel methods with different kernels. We empirically studied different combinations of feature rescaling, feature extraction, feature selection, algorithms, and numbers of features and qubits to assess where quantum kernels could provide advantages on our datasets. Our results revealed that quantum kernel methods achieve high classification performance, similar to classical ML methods, with some advantages in several cases. To achieve these results, a few critical parameters must be considered regarding data manipulation, one of which is feature engineering. Depending on the method that we selected for feature rescaling, the accuracy of the classification varied significantly. Another parameter was the method that we chose for reducing the dimension of features. In our case, reducing the data from 43 to 2 dimensions using feature extraction, provided a lower classification accuracy compared with classical algorithms. On the other hand, the use of feature selection methods to reduce the number of features to 5, 10, and 20 significantly improved the accuracy of the quantum models. Another crucial parameter to specify is a suitable feature map. The heuristic application of a ZZFeatureMap and the encoding function described by Havlíček et al.[4] provided better results than the others that we assessed in this study[8]. The quantum algorithm that trains a quantum kernel with QKA[16] also improved classification results. Finally, the size of the dataset is another parameter to consider. As in the classical world, we observed that the classification accuracy significantly varied according to the sample size.

The dataset we used was a compromise between achieving sufficient statistical power and the current limitations of quantum computing resources. As quantum technology advances, we anticipate being able to process larger datasets. Additionally, neuroscience itself lacks digitized neurons. The number of available neuronal morphologies for classification is currently limited[33–36]. This is a challenge for the neuroscience community. This limitation indeed calls for innovative techniques to increase the variety and volume of neuronal morphologies at our disposal for more comprehensive studies, which could be addressed by generative artificial intelligence. However, we believe that our current work offers valuable preliminary insights into the capabilities and potential of quantum machine learning in the context of neuron classification. In parallel, as the field of quantum technology continues to advance, we anticipate an increased capacity to process more extensive datasets. This casts light on the capabilities and prospective benefits of employing quantum machine learning for neuron classification, acting as a stepping stone for future, more expansive research in this burgeoning field.

Huang et al.[37] underscore that despite the potential advantages of quantum computing, we must concede that present-day quantum systems are beset with noise and other errors. Full-fledged, fault-tolerant quantum computers remain an aspirational goal. Another paper by Ezratty[38] astutely notes that even with significant advancements, we have yet to witness the implementation of a use case that resolves actual problems faster or more energy-efficiently than classical supercomputers. We operate within the constraints of the noise and small-scale while recognizing the importance of techniques and strategies highlighted by Ezratti. These include enhancing qubit fidelities, various types of quantum error mitigation methods, and the amalgamation of analog and digital approaches. We perceive them as active research areas with the potential to increase the reliability and robustness of our methodology. A recent paper published by Kim et al.[39] in Nature presents an experimental demonstration of the remarkable potential of quantum computers to surpass classical simulations by effectively leveraging error learning and mitigation techniques within the system. Their study successfully generated extensive entangled states, which were utilized to simulate the intricate dynamics of spins within a material model. The results yielded precise predictions of crucial properties such as magnetization. This notable study provides compelling evidence for the practical utility of quantum computing even in a pre-fault-tolerant era. As researchers in quantum machine learning, we maintain flexibility and readiness to evolve our methodologies as the field progresses. Methods such as variational quantum algorithms, error mitigation strategies, and quantum circuit compilation are being developed and refined to overcome challenges and maximize the utility of quantum devices. These methods, aimed at counteracting the noise and errors in quantum systems, form a crucial part of our quantum machine learning methodology. Within the context of our work, we have a particular interest in variational quantum algorithms. They allow us to design and optimize quantum circuits for better estimation of quantum kernels, even in the presence of noise. Additionally, error mitigation strategies help us to correct for quantum noise impacting our computation, thus improving the reliability of our results. Moreover, our work also involves benchmarking our quantum machine learning approach against classical algorithms to demonstrate its effectiveness.

Quantum ML can be used on real-world datasets and potentially lead to superior classification results compared with classical methods. Our primary objective was to explore the potential of quantum kernel algorithms in this particular domain of neuron morphology classification, investigate their feasibility and their impact on classification accuracy. By harnessing the capabilities of quantum systems, we seek to advance the field of neuroscience by introducing a novel approach that may offer distinct advantages over classical methods. The unique nature of neuron morphology classification presents its own set of challenges. Neuron classification is a complex task that requires capturing intricate structural details, and traditional machine learning approaches struggle to fully exploit the underlying patterns within the data. By introducing quantum kernel algorithms, we aim to harness the power of quantum computation to potentially enhance the accuracy and effectiveness of neuron morphology classification. Therefore, our contribution lies in the application and adaptation of quantum machine learning techniques to this specific problem.

Over the course of this study, we demonstrate a pathway that quantum ML can be applied to neuron morphology data. We believe that quantum ML will play a key role in the progress of neurosciences.

## Methods

**Datasets.** Morphology-based, whole rat brain neuron type classification is challenging, given the significant number of neuron types, limited reconstructed neuron samples, and diverse data formats. For our work on neuron morphology classification, we accessed NeuroMorpho.org, the largest collection of publicly accessible 3D neuronal reconstructions and one of the integrated resources in the Neuroscience Information Framework. It offers the opportunity to use morphological data in the context of other relevant resources and diverse subdomains of neuroscience[28]. The morphologies of neurons are usually described, stored, and shared digitally in two types of neuronal formats: SWC-format files (with data compressed but unstructured) and the 2D- or 3D-image format (structured but with high information dimensions)[24]. Digitally reconstructed neurons can be used and re-used in various research projects with different scientific aims[30].

One challenge during classification relate to the diverse data formats, such as 2D and 3D images (structured, with high dimensions and fewer samples than the complexity of morphologies) or SWC-format files (low dimensional and unstructured)[33]. We accessed digitally reconstructed neurons by species, brain regions, and cell types on NeuroMorpho.org[30]. The database contains 173,821 cells derived from/distributed over more than 300 regions. Tools already exist that allow researchers to extract quantitative morphological characteristics from neuronal reconstructions, such as L-Measure[40] and NeuroM (https://neurom.readthedocs.io/en/stable/index.html). The reconstructions are usually obtained from fluorescence microscopy preparations or brightfield, which can be synthesized through computation simulations. We extracted features from 27,881 labeled rat neurons. The data contained three main classes (principal, interneuron, and glial cells) with 14 subclasses, including six types of principal cells (ganglion, granule, medium spiny, parachromaffin, Purkinje, and pyramidal cells), six types of interneurons (basket, chandelier, Martinotti, double bouquet, bitufted, and nitrergic) and two types of glial cells (microglia and astrocytes; Fig. 1). Then, from the dataset, we extracted five samples (Fig. 1) for multiclass (ganglion, granule, medium spiny, parachromaffin, purkinje, pyramidal, basket, bitufted, chandelier, double bouquet, Martinotti, nitrergic, astrocytes, and microglia) classification. A total of 43 morphological features were extracted for each reconstructed neuron using the L-Measure tool[40] (Fig. 1). We then applied our algorithms on the five samples with a Mahalanobis transformation and the suppression of a neuron when the surface of the soma equaled 0 (Fig. 1). The Mahalanobis distance is the multivariate metric that measures the distance between a point and a distribution. Applying the Mahalonobis distance allowed us to reduce the standard deviation for each feature by deleting neurons from our dataset. The datasets were preprocessed to deal with missing values. If a value within the features was missing, the neuron was deleted from the dataset. Categorical features such as morphology types were encoded, transforming each categorical feature with m possible values.

**Data preprocessing.** To assess the impact of feature engineering techniques on classical and quantum algorithms, we ran combinations of feature rescaling, feature extraction, and feature selection. For both feature extraction and feature selection, the following feature rescaling techniques were applied:

- The StandardScaler, which removes the mean and scales to unit variance: $z = \frac{x - mean(x)}{standard\ deviation(x)}$
- The MinMaxScaler method: $z = \frac{x_i - \min(x)}{\max(x) - \min(x)}$
- The MaxAbsScaler method: $z = \frac{x_i}{\max(abs(x))}$
- The RobustScaler method: $z = \frac{x_i - Q_1(x)}{Q_3(x) - Q_1(x)}$
- The l2-normalization method: $y = ||x||_2 = \sqrt{\sum_{i=1}^{n} x_i^2}$
- Logistic data transformation: $z = \frac{1}{1 + e^{-x}}$
- Lognormal transformation: $F(x) = \phi\left(\frac{\ln(x)}{\sigma}\right), x \geq 0; \sigma \geq 0$
- Box–Cox (for all x strictly positive): $B(x, \lambda) = \begin{cases} \frac{x^\lambda - 1}{\lambda} & if\ \lambda \neq 0 \\ \log(x) & if\ \lambda = 0 \end{cases}$
- Yeo-Johnson: $\psi(\lambda, y) = \{$ ——
- The nonparametric quantile transformation (normal and uniform distribution), which transforms the data to a certain data distribution, such as normal or uniform distribution, by applying the quantile function, an inverse function of the cumulative distribution function, to the data. Let X be a random variable following a normal distribution:

$$X \sim \mathcal{N}(\mu, \sigma^2)$$

Then, the quantile function of X is

$$Q_X(p) = \sqrt{2\sigma} \cdot erf^{-1}(2p - 1) + \mu$$

where $erf^{-1}(x)$ is the inverse error function.

Let X be a random variable following a continuous uniform distribution:

$$X \sim \mathcal{U}(a, b)$$

Then, the quantile function of X is

$$Q_X(p) = \begin{cases} -\infty, & \text{if } p = 0 \\ bp + a(1-p), & \text{if } p > 0 \end{cases}$$

After rescaling the data with each of the aforementioned methods, we combined the outputs with feature extraction techniques, which led to more than 150 possible combinations per algorithm. For feature extraction, we reduced the dataset to two features (i.e., two principal components). The following feature extraction techniques were tested: PCA, ICA, ICA with PCA, LDA, random projection, truncated SVD, Isomap, neighborhood component analysis, t-distributed stochastic neighbor embedding (t-SNE), locally linear embedding (standard LLE), Hessian locally linear embedding (Hessian LLE), MLLE, LTSA, and multidimensional scaling.

Another way to reduce the dimensionality of the dataset is to experiment with feature selection techniques that assign a score to input features based on how useful they are for predicting a target variable. This also allows an enhanced understanding of which features are crucial for classifying neurons. In this study, we combined all of the feature scaling techniques with the following three tree-based feature selection embedded methods: random forest, decision tree, and XGboost. Such tree-based algorithms are commonly used for prediction. They can also be an alternative method for selecting features by revealing which features are more important, and which are the most used, in making predictions on the target variable (classification). Take random forest for example: It is an ML technique used to solve regression and classification and consists of many decision trees; each tree of the random forest can calculate the importance of a feature. The random forest algorithm can do this because of its ability to increase the pureness of the leaves. In other words, when one trains a tree, feature importance is determined as the decrease in node impurity weighted in a tree—the higher the increment in leaf purity, the more important the feature. It is called pure when the elements belong to a single class. After normalization, the sum of the importance scores calculated is 1. The mean decreased impurity is called the Gini index (between 0 and 1), which is used in random forest to estimate a feature's importance, measuring the degree or probability of a variable being wrongly classified when it is randomly chosen. The index is 0 when all elements belong to a certain class, 1 when the elements are randomly distributed across various classes, and 0.5 when the elements are equally distributed into some classes. The Gini index is calculated as follows:

$$Gini = 1 - \sum_{i=1}^{n}(p_i)^2$$

where $p_i$ is the probability of an element being classified to a distinct class.

**Kernel methods.** Classifying data using quantum algorithms could provide advantages, such as faster kernel computing compared with classical computing as well as improved classification accuracy. To achieve such advantages, finding and explicitly specifying a suitable feature map is essential; however, this is not straightforward compared with classic kernel specification. Although theoretical work has demonstrated advantages on synthetically generated datasets, concrete, specific applications are required to empirically study whether quantum advantage can be achieved and for what types of datasets and applications. The challenge lies in finding quantum kernels that could provide advantages on now real-world datasets.

In recent years, SVMs have been widely used as binary classifiers and applied to solve multiclass problems. In binary SVMs, the objective is to create a hyperplane that linearly divides n-dimensional data points into two components by searching for an optimal margin, which correctly segregates the data into different classes. The hyperplane that divides the input dataset into two groups can either be in the original feature space or in a higher dimensional kernel space. The selected optimal hyperplane among many hyperplanes that might classify the data corresponds to the one that has the largest margin and that allows the largest separation between classes. This is an optimization problem under constraints, where the distance between the nearest data point and the optimal hyperplane (on each side) is maximized. The hyperplane is then called the maximum-margin hyperplane, which allows one to create a maximum-margin classifier. The closest data points are known as support vectors, while the margin is an area that generally does not contain any data points. If the hyperplane defined as optimal is too close to data points and the margin is too small, predicting new data will be difficult and the model will fail to generalize well.

Studies have built multiclass SVMs based on a binary SVM, such as the all-pair approach, where a binary classification problem for each pair of classes is used[41]. In addition to linear classification, it is also possible to compute a nonlinear classification using what is commonly called the kernel trick (kernel function), which maps inputs into high-dimensional feature spaces. The kernel function corresponds to an inner product of vectors in a potentially high-dimensional Euclidian space, which is referred to as the feature space. The objective of a nonlinear SVM is to gain separation by mapping the data to a higher dimensional space, as many classification or regression problems are not linearly separable or regressable in the space of the inputs $x$. The aim is to use the kernel trick to move to a higher-dimensional feature space given a suitable mapping $x \to \phi(x)$.

Compared with classical ML, one of the present study's objectives is to find better data patterns within ML processes by leveraging quantum systems that map data to higher dimensions for training purposes and for use by the scientific community. In 2014, Rebentrost et al.[42] proposed a theoretically feasible quantum kernel approach based on SVM. In 2019, Havlíček et al.[4] as well as Schuld and Killoran[14] presented two mplementations of quantum kernel methods. Recently, Havlíček et al.[4] experimentally implemented two quantum algorithms in a superconducting processor. The quantum variational classifier, as in conventional SVM, uses a variational quantum circuit to classify data, while the quantum kernel estimator estimates the kernel function and optimizes the classical SVM. The principle, considering a classical data vector $x \in \chi$, is to map $x$ to a n-qubit quantum feature state $|\phi(x)\rangle$ by a unitary encoding circuit $U(x)$, such as $\phi(x) = U(x)|0^n 0^n|U^\dagger(x)$. For two samples $x, \widetilde{x}$, the quantum kernel function K[12] is defined as the inner products of two quantum feature states in the Hilbert–Schmidt space $K(\boldsymbol{x}, \tilde{\boldsymbol{x}}) = tr[\phi^{\dagger}(\boldsymbol{x})\phi(\tilde{\boldsymbol{x}})]$ and translates as the transition amplitude $K(\boldsymbol{x}, \tilde{\boldsymbol{x}}) = |0^n|U^{\dagger}(\boldsymbol{x})U(\tilde{\boldsymbol{x}})|0^n|^2$. For instance, the kernel function can be estimated on a quantum computer through a procedure called quantum kernel estimation (QKE), which consists of evolving the initial state $|0^n\rangle$ with $U^{\dagger}(\boldsymbol{x})U(\tilde{\boldsymbol{x}})$ and recording the frequency of the all-zero outcome $0^{n[16]}$. The constructed kernel is then injected into a standard SVM. By replacing the classical kernel function with QKE, it is possible to classify data in a quantum feature space with SVM.

*Quantum kernel method.* Although the principle is almost the same as the classical kernel method, the quantum kernel method is a kernel method based on quantum computing properties that maps the data point from an original space to a quantum Hilbert space. The quantum mapping function is critical in quantum kernel methods and has a direct impact on the model's performance. Finding a suitable feature map in the context of gate-based quantum computers is less trivial than just specifying a suitable kernel on classical algorithms. In quantum kernel ML, a classical feature vector $\vec{x}$ is mapped to a quantum Hilbert space using a quantum feature map $\Phi(\vec{x})$, such that $K_{ij} = |\Phi^{\dagger}(\vec{x}_j)|\Phi(\vec{x}_i)|^2$. There are crucial factors to evaluate when one considers a feature map, such as the feature map circuit depth, the data map function for encoding classical data, the quantum gate set, and the order expansion. One can also find different types of feature maps. One example is the ZFeatureMap, which implements a first-order diagonal expansion where $|S| = 1$. One must set up various parameters: (i) feature dimensions (the dimensionality of the data, which equals the number of required qubits); (ii) the number of times the feature map circuit is repeated (reps); and (iii) a function that encodes the classical data. Here, no entanglement occurs as there are no interactions between features. Another example is the ZZFeatureMap, which is a second-order Pauli-Z evolution circuit that allows $|S| \leq 2$ and $\Phi$ a classical nonlinear function. Here, interactions in the data are encoded in the feature map according to the connectivity graph and the classical data map. Similar to ZFeatureMap, ZZFeatureMap requires the same parameters as well as an additional one, namely the entanglement that generates connectivity ('full', 'linear', or own entanglement structure). The PauliFeatureMap is the general form that allows one to create feature maps using different gates. It transforms input data $\vec{x} \in \mathbb{R}^n$ as follows:

$$U_{\Phi(\vec{x})} = \exp\left(i \sum_{S \subseteq [n]} \phi_S(\vec{x}) \prod_{i \in S} P_i\right)$$

where $P_i \in \{I, X, Y, Z\}$ denotes the Pauli matrices, and S the connectivities between different qubits or datapoints: $S \in \{\binom{n}{k} combinations, k = 1, \ldots, n\}$. For k = 1 and $P_0$, we can refer to ZFeatureMap and ZZFeatureMap for k = 2 and $P_0 = Z$ and $P_{0,1} = ZZ$[4].

Eight algorithms were used in the present study. The first one we named q_kernel_zz, which applies a ZZFeatureMap. As described by Havlíček et al.,[4] we defined a feature map on n-qubits generated by the unitary:

$$\mathcal{U}_{\Phi}(\vec{x}) = U_{\Phi(\vec{x})} H^{\otimes n} U_{\Phi(\vec{x})} H^{\otimes n}$$

where H denotes the conventional Hadamard gate, and $U_{\phi(\vec{x})}$ is a diagonal gate in the Pauli-Z basis.

$$U_{\Phi(\vec{x})} = \exp\left(i \sum_{S \subseteq [n]} \phi_S(\vec{x}) \prod_{i \in S} Z_i\right)$$

Taking an example with two qubits, the general expression is as follows:

$$\mathcal{U}_{\Phi}(\vec{x}) = U_{\Phi(\vec{x})} H^{\otimes 2} U_{\Phi(\vec{x})} H^{\otimes 2}$$

where

$$U_{\Phi(\vec{x})} = \exp\left(i\phi_1(x)ZI + i\phi_2(x)IZ + i\phi_{1,2}(x)ZZ\right)$$

The encoding function that transforms the input data into a higher dimensional feature space was given by

$$\Phi(x) = \{\phi_1(x), \phi_2(x), \phi_{1,2}(x)\}$$

To create a feature map and test different encoding functions, we used the encoding function from Havlíček et al.[4] as follows:

$$\phi_{\{i\}}(x) = x_i \text{ and } \phi_{\{1,2\}}(x) = (\pi - x_1)(\pi - x_2) \quad (q\_kernel\_zz)$$

For q_kernel_zz, we used the ZZFeatureMap with full entanglement, a different feature dimension depending on the dimensionality reduction techniques (we used two qubits when using feature extraction techniques and five with feature selection techniques); moreover, we repeated the data encoding step two times.

The second algorithm that we tested was q_kernel_default, which applies a PauliFeatureMap (paulis = ['ZI','IZ','ZZ']) with the default data mapping $\phi_S$:

$$\phi_S(\vec{x}) = \begin{cases} x_0 & \text{if } k = 1 \\ \prod_{j \in S} (\Pi - x_j) & \text{otherwise} \end{cases}$$

For q_kernel_default, we used the PauliFeatureMap with full entanglement, a different feature dimension depending on the dimensionality reduction techniques, and we repeated the data encoding step two times.

In addition, we used the five encoding functions presented by Suzuki et al.,[8] which are presented as follows for a two-qubit example:

$$\phi_{\{i\}}(x) = x_i \text{ and } \phi_{\{1,2\}}(x) = \pi x_1 x_2 \quad (\text{q\_kernel\_8})$$

$$\phi_{\{i\}}(x) = x_i \text{ and } \phi_{\{1,2\}}(x) = \frac{\pi}{2}(1 - x_1)(1 - x_2) \quad (\text{q\_kernel\_9})$$

$$\phi_{\{i\}}(x) = x_i \text{ and } \phi_{\{1,2\}}(x) = \exp\left(\frac{|x_1 - x_2|^2}{8/\ln(\pi)}\right) \quad (\text{q\_kernel\_10})$$

$$\phi_{\{i\}}(x) = x_i \text{ and } \phi_{\{1,2\}}(x) = \frac{\pi}{3\cos(x_1)\cos(x_2)} \quad (\text{q\_kernel\_11})$$

$$\phi_{\{i\}}(x) = x_i \text{ and } \phi_{\{1,2\}}(x) = \pi \cos(x_1)\cos(x_2) \quad (\text{q\_kernel\_12})$$

For q_kernel_8, q_kernel_9, q_kernel_10, q_kernel_11, and q_kernel_12, we used the PauliFeatureMap (paulis = ['ZI','IZ','ZZ']) with full entanglement and repeated the data encoding step two times.

It is also possible to train a quantum kernel with QKA,[16] which iteratively adapts a parametrized quantum kernel to a dataset, converging to the maximum SVM margin at the same time. We named this q_kernel_training. The algorithm introduced by Glick et al.[16] allows a quantum kernel to be learned from a family of kernels (covariant quantum kernels that are related to covariant quantum measurements) and simultaneously converge to the maximum SVM margin, optimizing the parameters in a quantum circuit. To implement it, we prepared the dataset as mentioned previously and defined the quantum feature map. Then, we used the QuantumKernelTrained.fit method to train the kernel parameters and pass them to an ML model. In covariant quantum kernels, the feature map is defined by a unitary representation D($\boldsymbol{x}$) for $\boldsymbol{x} \in \chi$ and a state $|\psi\rangle = U|0^n\rangle$. The kernel matrix is given as follows[43]:

$$K(\boldsymbol{x}, \tilde{\boldsymbol{x}}) = \left|\langle 0^n | U^\dagger D^\dagger(\boldsymbol{x}) D(\tilde{\boldsymbol{x}}) U | 0^n \rangle\right|^2$$

For a given group, QKA is used to find the optimal fiducial state.[16] In the context of covariant quantum kernels, the equation can be extended to the following:

$$K_\lambda(\boldsymbol{x}, \tilde{\boldsymbol{x}}) = \left|\langle 0^n | U_\lambda^\dagger D^\dagger(\boldsymbol{x}) D(\tilde{\boldsymbol{x}}) U_\lambda | 0^n \rangle\right|^2$$

where quantum kernel alignment learns an optimal fiducial state parametrized by $\lambda$ for a given group.

To train the quantum kernel, we used QuantumKernel, holding the feature map and its parameters. For all of the trained quantum kernels, we passed them to an ML model (fit and test) using the Qiskit's QVC for classification (https://qiskit.org). For q_kernel_training, we used QuantumKernel and QuantumKernelTrainer to manage the training process. We selected the kernel loss function (SVCLoss) as input for the QuantumKernelTrainer. Furthermore, we selected the simultaneous perturbation stochastic approximation as the optimizer with 10 as the maximum number of iterations, 0.05 as the learning rate, and 0.05 for the magnitude of the perturbation for the finite difference approximation of the gradients.

*Support vector machine.* For comparison purposes, we used classical SVMs to classify neuron morphologies. We used the "kernel trick" to move to a higher-dimensional feature space given a suitable mapping $\boldsymbol{x} \to \phi(\boldsymbol{x})$. In the present study, we implemented SVMs with the following kernels:

- Linear: $K(x_i, x_j) = x_i^T x_j$
- Gaussian radial basis function: $K(x_i, x_j) = e^{-\frac{\|x_i - x_j\|^2}{2\sigma^2}}$
- Polynomial: $K(x_i, x_j) = (\boldsymbol{x}_i \cdot \boldsymbol{x}_j + a)^b$
- Sigmoidal: $K(x_i, x_j) = tanh(a\boldsymbol{x}_i \cdot \boldsymbol{x}_j - b)$

The scikit-learn svm.SVC algorithm (http://scikit-learn.org) was also used. All algorithms were run with all of the different combinations of feature rescaling/feature extraction and feature rescaling/feature selection techniques.

**Quantum simulator.** The quantum simulations were run on StatevectorSimulator, which is available on Qiskit and IBM Quantum Services. Each circuit was run with 1024 shots. The statevector_simulator was run on Nvidia GPUs.

**Classical hardware.** We ran the algorithms on a computer with a 12-core (24-thread) AMD Ryzen 9 3900X 12-Core Processor, 32 GB of memory, and Nvidia GeForce RTX 3070 (Driver version: 510.85.02, CUDA version: 11.6).

**Quantum hardware.** The following four 27-qubit superconducting quantum computers available on IBM Quantum Services were used to run the quantum algorithms: ibmq_kolkota, ibmq_montreal, ibmq_mumbai, and ibmq_auckland. The qubit connectivity and the system characteristics are provided in Fig. 3.

**Software.** The algorithms were implanted in Python 3 (http://www.python.org) using the scikit-learn (https://scikit-learn.org/stable/) open-source python library. Morphological features were extracted for each neuron using the L-Measure tool,[40] allowing the extraction of quantitative morphological measurements from neural reconstructions. The open-source SDK Qiskit (https://qiskit.org) was used to work with quantum algorithms. Based on the aim of the present study, we developed an open-source framework that allowed us to run all of the pipelines; it is available on GitHub: https://github.com/xaviervasques/hephaistos.

**Classification assessment and selection.** For each experiment, to assess whether patterns were identifiable, we trained the supervised classification algorithms using 80% of the data sample described in Fig. 1, which was randomly chosen, and we assessed the accuracy of predicting the remaining 20%. This method provided an indication of how well algorithms perform on data used for training; however, overfitting and underfitting may have occurred, thereby limiting our understanding of how well the learner would generalize to independent and unseen data. To optimally fit the model, we also applied the k-fold cross-validation test to better estimate accuracy. We computed the scores five consecutive times. The mean score and the 95% confidence interval are presented. Supervised algorithms were tested on neurons gathered by m-types in young and adult rat populations. We selected the top five algorithms where we applied feature extraction techniques and where data were dimensionally reduced to two features. The algorithms were chosen according to the classification accuracy scores based on k-fold cross-validation for both classical and quantum simulation algorithms (two qubits) applied to sample 5 (Fig. 1). To understand the behavior of the chosen algorithms, we then compared the results with the accuracy of the selected algorithms applied to other samples (Samples 1, 2, 3, and 4). In addition, we selected the top five algorithms where we applied feature selection techniques to use the five most important features. The algorithms were chosen according to the classification accuracy scores, which were based on k-fold cross-validation scores for both classical and quantum algorithms (five qubits) applied to sample 5 (Fig. 1). Finally, the quantum simulation algorithms that provided the highest accuracy were run on real quantum hardware on samples 1, 2, 3, and 4, allowing us to run the algorithms in a reasonable computing time.

## Data availability
The data that support the findings of this study are available from the corresponding author, X.V., upon reasonable request.

### Acknowledgements
We would like to thank Anna Lelorieux for her support on the data preparation phase.

### Author contributions
All co-authors have contributed to the concept design and/or drafting of the manuscript. All authors have read and approved the manuscript.

### Competing interests
The authors declare no competing interests.

### Additional information
**Correspondence** and requests for materials should be addressed to X.V.

**Reprints and permissions information** is available at www.nature.com/reprints.

**Publisher's note** Springer Nature remains neutral with regard to jurisdictional claims in published maps and institutional affiliations.

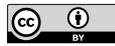 **Open Access** This article is licensed under a Creative Commons Attribution 4.0 International License, which permits use, sharing, adaptation, distribution and reproduction in any medium or format, as long as you give appropriate credit to the original author(s) and the source, provide a link to the Creative Commons licence, and indicate if changes were made. The images or other third party material in this article are included in the article's Creative Commons licence, unless indicated otherwise in a credit line to the material. If material is not included in the article's Creative Commons licence and your intended use is not permitted by statutory regulation or exceeds the permitted use, you will need to obtain permission directly from the copyright holder. To view a copy of this licence, visit http://creativecommons.org/licenses/by/4.0/.

© The Author(s) 2023